\documentclass[wias]{iosart2c}     

\usepackage[T1]{fontenc}
\usepackage{times}%
\usepackage{graphicx} 
 \usepackage{balance}
 \usepackage{flushend}
 \usepackage{multirow}
 \usepackage{mathtools}
 
 \usepackage{color,soul}
 \usepackage{cite}
 \usepackage{enumerate}

\usepackage{verbatim}
\usepackage{subfig}

\usepackage{algorithm}
\usepackage{algpseudocode}

\pubyear{0000}
\volume{0}
\firstpage{1}
\lastpage{1}

\begin{document}

\begin{frontmatter}    

\title{Memristive System Design for Variable Pixel G-Neighbor Denoising Filter}
\runningtitle{Memristive System Design for Variable Pixel G-Neighbor Denoising Filter}

\maketitle


\author[A]{Kamilla Aliakhmet\fnms{} \snm{}},
\author[A]{Diana Sadykova\fnms{} \snm{}},
\author[B]{Joshin Mathew\fnms{} \snm{}},
\author[A]{Alex Pappachen James\fnms{} \snm{}\thanks{Email:apj@ieee.org}}
\runningauthor{K. Aliakhmet,D. Sadykova,J. Mathew, A. P. James }
\address[A]{School of Engineering, Nazarbayev University}
\address[B]{ARS  Traffic \& Transport Technology, India}

\begin{abstract}
Image blurring artifact is the main challenge to any spatial, denoising filters. This artifact is contributed by the heterogeneous intensities within the given neighborhood or window of fixed size. Selection of most similar intensities (G-Neighbors) helps to adapt the window shape which is of edge-aware nature and subsequently reduce this blurring artifact. The paper presents a memristive circuit design to implement this variable pixel G-Neighbor filter. The memristive circuits exhibits parallel processing capabilities (near real-time) and neuromorphic architectures. The proposed design is demonstrated as simulations of both algorithm (MATLAB) and circuit (SPICE). Circuit design is evaluated for various parameters such as processing time, fabrication area used, and power consumption. Denoising performance is demonstrated using image quality metrics such as peak signal-to-noise ratio (PSNR), mean square error (MSE), and structural similarity index measure (SSIM). Combining adaptive filtering method with mean filter resulted in average improvement of MSE to about 65\% reduction, increase of PSNR and SSIM to nearly 18\% and 12\% correspondingly.

\end{abstract}

\begin{keyword}
Denoising filter\sep  G-neighbor\sep  memristor\sep  SPICE. 
\end{keyword}

\end{frontmatter}


\section{Introduction}
Image denoising methods are essentially representatives of low pass filtering where the noisy intra-region variations (high frequency component) is averaged-out and region-wise uniform intensities (low frequency component) \cite{BuadesCM05}. The simplest form of denoising filter is mean filtering where the denoised signal/intensity is the average value of neighboring intensities for a given fixed window size. In this approach the neighboring pixel locations are given uniform weights for averaging. Hence for neighborhoods with heterogeneous regions present, the denoised intensities assumes value ranging between mean intensities of individual regions within the selected window. This smooth variation between regions resulted in denoising is referred as blurring artifact. \par

Different varieties of edge-aware filtering techniques are thus introduced to address this artifact and produce images of best visual quality. Such filtering techniques are implemented to minimize smoothing operation over edge regions and maximize smoothing within individual regions. Few examples of such methods include Anisotropic Diffusion Filter, Bilateral Filter, Non-local Means Filter, Domain Transform Filter, Guided Filter, and $L_{0}$ (Gradient Minimization) Smoothing \cite{perona, 710815,1467423,GastalOliveira2011DomainTransform, 6319316,l0smoothing2011}. These examples give insight to fundamentals of image denoising algorithms. All these methods are presented as post-processing after image formation to enhance its visual quality. \par 

G-Neighbor filter presented a simplest form of on-hardware denoising implementation. Originally, G-Neighbors are the immediate connected neighboring pixels with respect to the reference location. The model calculates the similarity between pixels and only those neighboring locations with computed similarity greater than a predefined threshold are considered for averaging to estimate denoised value \cite{Gneib}. In this paper, we present an extended version of G-Neighbors and the same is implemented on hardware. \par

Hardware design of proposed variable pixel G-Neighbors are implemented using memristive circuits. Memristor is a nanoscale device that can demonstrate neuromorphic characteristics of biological neurons such as memory storage, logical operations (gates) and mathematical operations such as addition, subtraction, multiplication and Fourier Transformation \cite{22}. This technology is thus used as a component for the implementations of neuromorphic circuits capable of various image processing techniques such as edge detection and feature descriptor computations \cite{pajouhi2016image,7300445}. This paper presents the denoising implementation on hardware. \par

The design is evaluated for hardware performance parameters such as processing time, fabrication area usage and power consumption. Another part of evaluation includes the qualitative and quantitative performance of proposed variable pixel G-Neighbor filter. This  part evaluates the aspects such as noise reduction (Mean Square Error, MSE), signal strength enhancement (Peak Signal-to-Noise Ratio, PSNR), and amount of structural preservation (Structural Similarity Index Measure, SSIM) \cite{Gonzalez02a,1284395}. \par

\section{Methodology}
\subsection{Variable Pixel G-Neighbor Filter}
G-Neighbors are the most similar pixels out of either 4 or 8 connected neighbors for a 3 $\times$ 3 window/neighborhood size. The similarity is computed as a function of gray-level distance between two pixel intensities and similarities above a given threshold define the G-Neighbor. The parallel architecture implemented using G-Neighbor was referred as pipelined image processing engine (PIPE). In this architecture, there is a video stage where the input intensity signal is digitized and further modular processing stage where the neighbor operation is performed. \par  

\begin{figure}[tbh]
\begin{centering}
\includegraphics[width=3in]{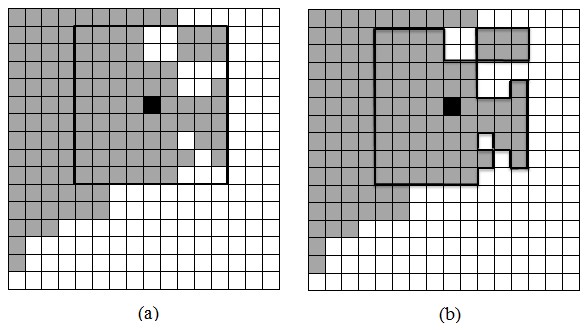}\\
\par \end{centering}
\caption{Filtering mask produced for (a) conventional and (b) variable pixel filtering methods}
\label{fig:WindowSelection}
\end{figure}

In this paper, this algorithm is further extended and modified to match with functional requirements with higher window sizes. Instead of considering only 4 or 8 connected neighbors, the modified version selects all similar pixels within a given neighborhood/window size ($w \times w$). As the method selects only similar pixels, the neighborhood/window adapts the shape of parent region (reference pixel's region) within it. This helps in avoiding pixels from heterogeneous regions from smoothing/averaging operation and converge as an edge-aware denoising filter( Fig. \ref{fig:WindowSelection}).  

The denoisied intensity at $(i, j)^{th}$ image location for a given noisy image ($I_{noisy}$) of size $M \times N$ is defined as in Eq. \ref{Eq::DenoisedIntensity}.

\begin{equation}
I_{(i,j)} =  \frac{\sum\limits_{s=-\frac{(w - 1)}{2}}^{\frac{(w - 1)}{2}}\sum\limits_{t=-\frac{(w - 1)}{2}}^{\frac{(w - 1)}{2}} S_{(s,t)} I_{(i+s,j+t)}}{w^2}
\label{Eq::DenoisedIntensity}
\end{equation}

Function $S(.)$ obtains the weight deciding whether a neighboring pixel is a G-Neighbor or not. Weight assumes values $0$ and $1$, where $1$ represents the G-Neighbor. The weight calculation function is defined as in Eq. \ref{Eq::WghtCalcFun}

\begin{equation}
    S_{(s,t)} = 
\begin{cases}
    1, &\text{if}\;\;\;|I_{(i,j)}-I_{(i+s,j+t)}|\leq\eta \\
    0,              & \text{otherwise}
\end{cases}\,,
\label{Eq::WghtCalcFun}
\end{equation}

Eq. \ref{Eq::WghtCalcFun} used grayscale difference/distance between pixel intensities as the similarity measure. We can also use other similarity metrics to measure inter-pixel closeness and Shepherd's similarity measure is one of such example. However, for the easy of circuit implementation of inter-pixel similarity calculation follows Eq. \ref{Eq::WghtCalcFun}. Also the algorithm is demonstrated in Algorithm 1.

\begin{algorithm}
\label{Algo::GNeighborFilter}
   \caption{Variable Pixel G-Neighbor Filter}
    \begin{algorithmic}[1]
      \Function{GNeighborFilter}{$Img,w,\eta$}\Comment{Where Img - input image, w - window size, $\eta$ - similarity threshold}

      \State ${w_{r}} = (w-1)/2$\Comment{window/neighborhood size radius}
      \State ${Img_{Zpad}} = zeroPadArray(Img, [w_{r},w_{r}])$\Comment{adding zeros to image peripherals to address indexing issues at borders}
      \State ${[rows,cols]} = size(Img)$\Comment{finding image size}
      \State $F = zeros$([$rows$, $cols$])\Comment{output image is initialized as a zero image}
      \For{${r=w_{r}}$ to ${rows-w_{r}}$}
        \For{${c=w_{r}}$ to ${cols-w_{r}}$}
        	\State $window = Img_{Zpad}$($r-w_{r}$ to $r+w_{r}$, $c-w_{r}$ to $c+w_{r}$)\Comment{window/neighborhood selection}
            \State $DImg = |window - window(w_{r}+1, w_{r}+1)|$ \Comment {difference/distance image}
            \State $windowMask = DImg \leq \eta$\Comment {Variable Pixel G-Neighbors are those with value$1$}
            \State $NZcount = countNonZeros(windowMask)$\Comment counting G-Neighbors
            \State $F(r-w_{r}, c-w_{r}) = sum(window.*windowMask)/NZcount$ \Comment {pixel-to-pixel masking and output calculation}
          \EndFor
      \EndFor
      \EndFunction
\end{algorithmic}
\end{algorithm}

\subsection{Neuromorphic circuits using Memristors}
Neuromorphic circuits are inspired from biological neurons, the smallest unit of brain implementing decision making and information storage \cite{ibrayevdesign,maan2016survey,maan2015memristor, olga2017htm, fedorova2016htm}. Memristive logic circuits are helpful in implementing hardware replica of a neuron. Such an implementation is capable of both data storage and logical gate operations. \par

Memristor is a nanoscale device that can be configured into ultradense crossbar by placing insulating film between two layers of nanowire electrodes so that two-terminal resistive switching device is formed at the crossing points \cite{7}. The latest developments using memristor crossbar architectures include non-volatile random access memory or NVRAM \cite{9}, configurable logic cells \cite{10}, and neuromorphic architectures \cite{11}. Image processing systems have been also proposed relying on the nonlinear and variable characteristics of the memristor \cite{12,14}.\par

Inspired by neuronal firing and training mechanisms in the human brain, threshold logic is used to design various computational blocks, including logic gates, multiple multiplication and addition, and Fourier Transform (FT) \cite{19,21}. Due to high-density packaging and switching characteristics, memristive threshold logic (MTL) circuits outperform conventional CMOS-based TL topologies in terms of low chip area and low leakage current \cite{22}. 

In this paper, memristive circuit is used to implement XOR gates to perform the pixel-to-pixel comparison leading to distance calculation. Such computed distance is further checked to identify G-Neighbors for smoothing operation.

\subsection{Proposed Design}
The focus of this paper is to propose an analog design for variable pixel G-Neighbor filter using memristive circuits.Hardware version of VPGNF should include following stages such as,
\begin{enumerate}
\item storing image to a hardware device
\item distance computation between reference pixel and respective neighbors (at individual pixel locations within defined neighborhood)
\item a threshold operation to identify G-Neighbors (at individual pixel locations within defined neighborhood)
\item an averaging operation to computed denoisied pixel value (at individual pixel locations within defined neighborhood)
\item storing the denoised image to a hardware device
\end{enumerate}

\begin{figure*}[tbh]
\centering
\includegraphics[width=6in]{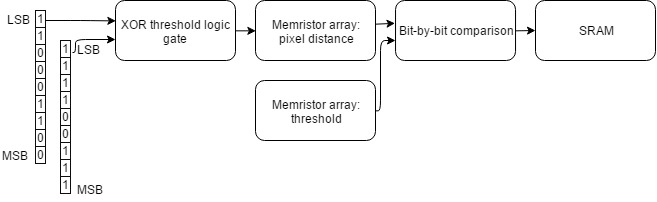}\\
\caption{An abstract view on hardware realization of Variable Pixel G-Neighbor Filter}
\label{fig::SystemOverview}
\end{figure*}

An abstract view on the design at individual pixel location is demonstrated in Fig. \ref{fig::SystemOverview} 

At each pixel locations, depending upon the given neighborhood/window sizes the reference pixel (8-bit grayscale) is initially compared against a neighbor (8-bit grayscale) and the respective distance/difference is calculated using an XOR threshold logic gate (XOR-TLG). The calculated distance is then stored to first-level memristor cross-bar array. Along with the first-level array, there is another memristor array with pre-stored threshold values, which is further used to identify G-Neighbor. These two arrays are the input to next stage which is the bit-by-bit comparison identifying the G-Neighbors and the same to stored to the static random access memory (SRAM) memory block and such identified neighbors are further considered for denoising stages. Note that in all stages of the proposed schematic, the binary states (bits) $0$ and $1$ are implemented as voltage levels $V_L$ and $V_H$ respectively. \par

Technical descriptions on each blocks in the schematic are added in below subsections.

\subsubsection{XOR Threshold Logic Gate (XOR-TLG)}
\begin{figure}[tbh]
\begin{centering}
\subfloat[\label{fig::MTL}]{\centering{}
\includegraphics[width=2in]{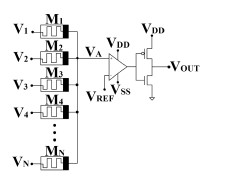}
}\quad
\subfloat[\label{fig:sim1}]{\centering{}\includegraphics[width=3in]{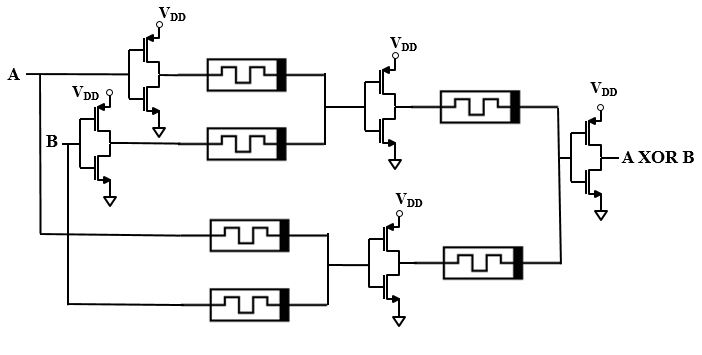}}
\par\end{centering}
\protect\caption{ (a) N-input MTL cell proposed by \cite{20} and (b) XOR threshold logic gate implemented using this cell\label{fig::mtlcell}}
\label{fig::XORSem}
\end{figure}

XOR threshold logic gate used is of two inputs and is constructed using MTL cells. They are used in create NOR and NAND gates. Such NOR and NAND along  with boolean arithmetic logic is realized to the two input XOR gate. N-input MTL cell (Fig. \ref{fig::MTL}) and XOR-TLG schematic (Fig. \ref{fig::mtlcell}b) are presented in  \ref{fig::XORSem}.  

N-input MTL cell is proposed in \cite{20}. In this model, memristors are used to provide programmable synaptic weight to inputs and a CMOS inverter implement the threshold device. $V_A$ represents the average voltage of all inputs and the same cell is used to compute either NOR or NAND gates by controlling the $V_{REF}$ voltage of operational amplifier. In the proposed schematic, to minimize the overall chip (fabrication) area usage, operational amplifier is avoided in the design. Truth table for a two input MTL cell is presented in Table \ref{truthtable}, where threshold of the inverter is less than $V_{t1}$ and greater than ($V_H$ +$V_L$)/2 for NOR and NAND operations, respectively. 
\begin{table}[h]
\caption{Truth table for 2-input MTL cell}
\label{truthtable}
\begin{center}
    \begin{tabular}{lllll}
    \hline 
    $V_1$ & $V_2$  & $V_A$ & $NOR$  & $NAND$ \\ \hline
   $V_L$  & $V_L$  & ($V_L$ +$V_L$ )/2 & $V_H$ & $V_H$ \\ \hline
   $V_L$  & $V_H$  & ($V_L$ +$V_H$ )/2 & $V_L$ & $V_H$\\ \hline
   $V_H$  & $V_L$  & ($V_H$ +$V_L$ )/2 & $V_L$ & $V_H$\\ \hline
   $V_H$  & $V_H$  & ($V_H$ +$V_H$ )/2 & $V_L$ & $V_L$  \\ \hline
     \end{tabular}
\end{center}\par
\end{table}

\subsubsection{Memristor array}
\begin{figure}[tbh]
\begin{centering}
\subfloat[\label{fig:sim}]{\centering{}
\includegraphics[width=2in]{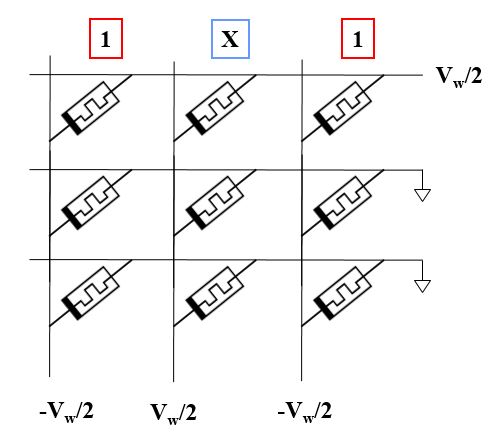}

}\quad \subfloat[\label{fig:sim1}]{\centering{}\includegraphics[width=2in]{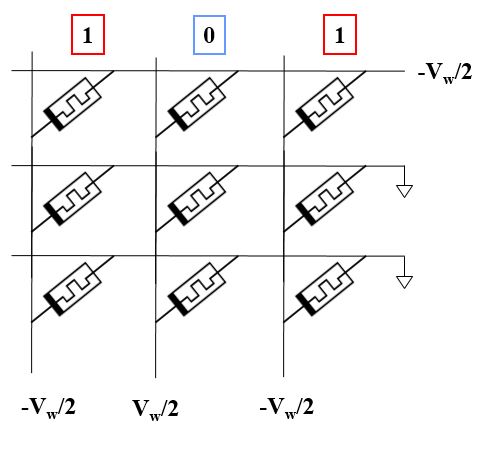}}
\par\end{centering}

\protect\caption{ Two stage writing process of bit values (a) '1' and (b) '0'\label{fig:Write10}}
\end{figure}

Both the distance between pixels and thresholding coefficient G are stored in two individual memristor arrays in two consecutive cycles (Fig. \ref{fig:Write10}). To write bit value of '1' (low resistance), $V_w$ larger than threshold voltage $V_{thr}$ for memristor switching ($V_w$\textgreater$V_{thr}$) must be applied across its terminals. On the other hand, -$V_w$ must be given to switch to state '0' (high resistance). In the first cycle, $V_w/2$ is applied on positive terminal, which means that columns with -$V_w/2$ applied will change to low resistance state, whereas ones with column voltage of $V_w/2$ will remain idle. When row voltage changes to -$V_w/2$, junctions with negative voltage difference across the device will switch to high resistance state. Most importantly, in the second cycle, values that were written in the previous cycle remain the same. \par
\begin{figure}[h]
\centering
\includegraphics[width=2in, height=2.5in]{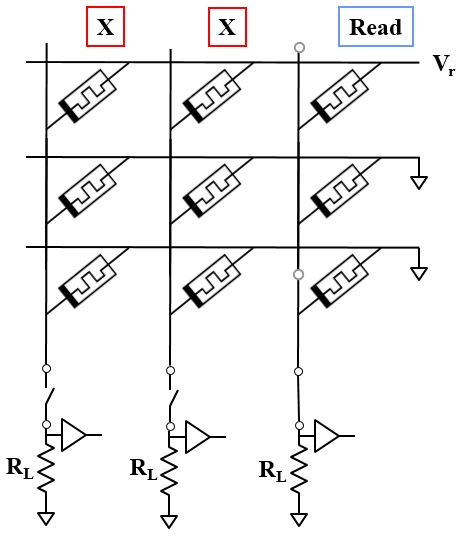}\\
\caption{Element-by-element reading process in the memristor array }
\label{fig:read}
\end{figure}

Read operation is performed by applying $V_r$, where $V_r$ \textless $V_{thr}$ to not to disturb state of the memristors (Fig. \ref{fig:read}). Voltage at the node of load resistor $R_L$ is detected column by column using inverter with threshold $V_{t2}$. Since rows are also addressed one by one and columns are isolated with switch, read operation becomes element-wise. 

\subsubsection{Bit-by-bit comparison}
\begin{figure}[tbh]
\centering
\includegraphics[width=3in, height=2in]{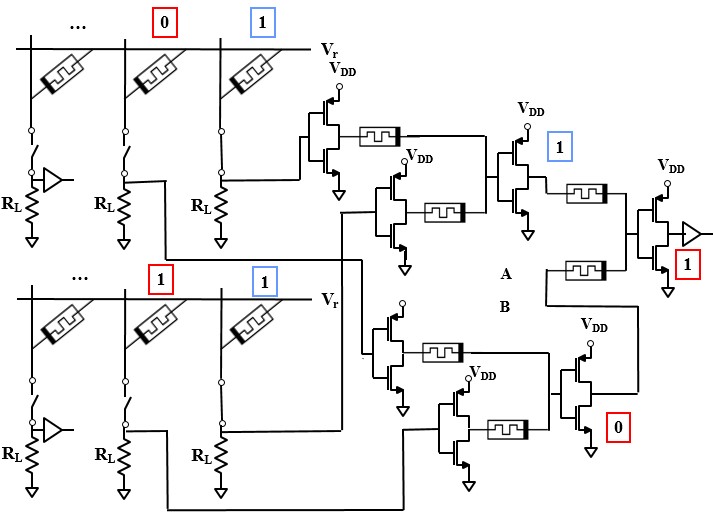}\\
\caption{Bit-by-bit comparison using memristor threshold logic }
\label{fig:bitbybit}
\end{figure}

This section describes configuration for bit-by-bit comparison which determines relation between two input pixels. After write operation is complete for all memristors, read pulses are applied and logic operations AND and OR are performed on the stored bit values using previously discussed logic cells. In Fig. \ref{fig:bitbybit}, schematic for comparison of last two bits of pixel distance and threshold coefficient is shown and values stored are '01' and '11'. The resultant bit '1' is saved in the memory block.

\subsubsection{Static Random Access Memory (SRAM)}
\begin{figure}[tbh]
\centering
\includegraphics[width=2.5in]{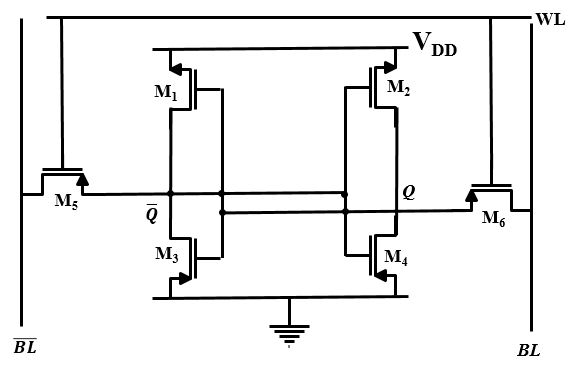}\\
\caption{ 1-bit Static Random Access Memory (SRAM) cell }
\label{fig:SRAMcircuit}
\end{figure}

Static Random Access Memory (SRAM) cell is commonly represented by 6T (transistor) configuration consisting of two cross-coupled inverters and two NMOS switches \cite{23} (Fig. \ref{fig:SRAMcircuit}). Data is stored and retrieved using word and bit line signals. The cell is on idle mode if word line is not active, and writing and reading operations can only be performed when it is asserted. 

\section{Image Quality Metrics}
The image quality metrics can be divided into subjective and objective methodologies. The first ones are of two types: absolute and relative grading, meaning, for instance, the best and the best in this room, correspondingly. This kind of assessment is entirely based on a visual image quality i.e. it does not have certain reliable criteria and mathematically not realizable \cite{zhang14}. Contrary, the second type is based on a numerical comparison with a reference image and set standards. In this section, objective methods will be described in terms of their methods of estimation and functions. \par

Objective image quality assessment method has three categories based on utilization of an original image. The first one is a full reference approach, which operates using both original and distorted images. It is considered to be the most accurate among these three objective methods, as it uses the whole source of information provided. The second one is a reduced reference mechanism, where original and noisy images are partially compared. Such type of evaluation can be practically used, as it can be easily embedded into a system \cite{carnec5}. Finally, the third one is a no reference method, which purely operates with a distorted image to produce results, meaning that its computation involves service information \cite{lan15}. Thus, in this paper, only full reference and reduced reference methods will be considered due to their effectiveness and wide field of applications. \par

The process of reference evaluation usually includes the use of MSE, PSNR, but their results can differ from human visual perception \cite{zhang14}. This happens as they simply represent the dissimilarity of pixels intensities. Therefore, the SSIM metric will be applied, as it performs image assessment more close to natural vision \cite{lan15}. However, some points of the image are not fully covered by SSIM evaluation, thus this paper will combine results from all methods: MSE, PSNR, and SSIM. \par

The first one is MSE, which is used to evaluate a number of deviations or image variance and it is calculated by taking an average of the square of the errors between processed and reference images \cite{huynh16}. In Eq. \ref{Eq::MSE}, M and N correspond to image size, while $I_1(s,t)$ and $I_2(s,t)$ correspond to location of image pixel intensities.

\begin{equation}
MSE=\frac{1}{MN} \sum_{s=1}^{M} \sum_{t=1}^{N} (I_1(s,t)-I_2(s,t))^2
\label{Eq::MSE}
\end{equation}

Another measure, which depends on MSE estimation, is PSNR. It is used to either evaluate enhancement of signal power or reconstruction losses \cite{wang17}. It is calculated by taking 10 times logarithm of maximum squared intensity to MSE ratio (see Eq. \ref{Eq::PSNR}).

\begin{equation}
PSNR = 10log (\frac{max(I)^2}{MSE}) 
\label{Eq::PSNR}
\end{equation}

It can be seen that PSNR increases when MSE value tends to reach zero, expressing higher image quality. On the other hand, low PSNR value means the high variation of compared image intensities.

The improved metric to simultaneously compare noise reduction and structural similarity preservation is SSIM \cite{md18}. It combines comparison of luminance (l), contrast (c) and structure (s). This measure assumes that reference image is of a good quality.

\begin{equation}
SSIM= function (l (I_1,I_2),c (I_1,I_2),s (I_1,I_2)) 
\label{Eq::SSIM}
\end{equation}

It can be seen from Eq. \ref{Eq::SSIM2} that luminance function estimates mean luminance to find their neighborhood. The construct and structure functions use standard deviation values in order to estimate closeness. 

\begin{equation}
\begin{cases}
	l (I_1,I_2)=\frac{2\mu_{I1}\mu_{I2}+c_1}{\mu_{I1}+\mu_{I2}+c_1} \\
	c (I_1,I_2)=\frac{2\sigma_{I1}\sigma_{I2}+c_2}{\sigma_{I1}+\sigma_{I2}+c_2} \\
    	s (I_1,I_2)=\frac{\sigma_{I1 I2}+c_3}{\sigma_{I1}\sigma_{I2}+c_3} \\
\end{cases}\,
\label{Eq::SSIM2}
\end{equation}

Additionally, the relationship between PSNR and SSIM in gray scale image derived by Hore and Ziou can be observed in Eq. \ref{Eq::SSIM3} \cite{hore4}.

\begin{equation}
PSNR= 10log_{10}\left[\frac{255^2}{2\sigma_{I1 I2}}\right]+10log_{10}\left[\frac{SSIM}{1-SSIM}\right] 
\label{Eq::SSIM3}
\end{equation}

From Eq. \ref{Eq::SSIM3} it can be seen that PSNR and SSIM are in some way dependent on each other. This feature is helpful for faster identification and prediction of threshold value.\par

The effectiveness of the algorithm in noise reduction was tested by adding salt and pepper noise. This noise is represented by randomly adding white and black pixels to an image with given frequency \cite{olsen19}.\par

This type of noise is usually removed by applying median filter operations \cite{chan20}. However, as it was mentioned before mean filter application has several advantages and may result in efficient denoising in adaptive filtering. Therefore both mean and median filter operations will be tested in the simulation.\par

In this paper, the image quality is evaluated in two ways; 1) Full reference and 2) Reduced reference evaluation.

\subsection{Full reference evaluation}
The evaluation metrics are applied to full images in this approach. The same metrics such as MSE, PSNR, and SSIM are evaluated. This provides a holistic understanding of the improvements made in terms of noise reduction, signal strength, and detail/structure preservation (see algorithm \ref{alg:fr}).\par

\begin{algorithm}
\caption{Full reference image quality assessment}\label{alg:fr}
\begin{algorithmic}[1]
\Procedure{FR}{$I$}
\State Read input image I.
\State N=mat2gray(I) \Comment{normalized between 0 and 1}
\State $[x,y]\gets $Size of image N
\State $N_p\gets $Padded image created from matrix N
\State \textbf{for} i $\gets$ 2 to x+1 \textbf{do}
\State 		\qquad \textbf{for} j $\gets$ 2 to y+1 \textbf{do}
\State \qquad\qquad $W=N_p (j-1:j+1,i-1:i+1)$\Comment{Neighborhood within 3x3 window}
\State \qquad\qquad $I_{difference}=|W-F(i,j)|$
\State \qquad\qquad $Neigh_G = I_{difference}$ \Comment{memristive similarity estimation} 
\State \qquad\qquad $Mask=Neigh_G<G$\Comment{similarity threshold for adaptive mask formation}
\State \qquad\qquad $Filter=(N.*Mask)$ \Comment{Filter application for adaptive window}
\State 		\qquad\textbf{end for}
\State \textbf{end for} 
\State \textbf{end procedure} 
\EndProcedure
\end{algorithmic}
\end{algorithm}

\subsection{Reduced reference evaluation}
In the reduced reference approach, only some part of original and distorted images is adopted for assessment. The reference part in this paper is chosen based on human visual perception. The human visual system reflects with different sensitivity to diverse structural data. Moreover, human eyes pay more attention to the regions, containing more information. Therefore variance can be used as a measure to detect regions of human visual perception \cite{lan15}. The square or circle regions of interest with more structural information are extracted by the algorithm \ref{alg:rrf} with following steps:
\begin{itemize}
\item	Read input image and normalize it between 0 and 1 in order to decrease illumination differences effect and make it faster for processing operations.
\item	Initialize logical region of interest
\item	The pixels, which located outside the region of interest, are assigned to zero.
\item	Finally, the reduced images are formed and ready to use.
\end{itemize}

\begin{algorithm}
\caption{Reduced reference image quality assessment}\label{alg:rrf}
\begin{algorithmic}[1]
\Procedure{RR}{$I$}
\State Read original input image I.
\State N=mat2gray(I) \Comment{normalized between 0 and 1}
\State $[row, column]\gets $Size of image N
\State Set circuit coordinates A, B and Radius, or square coordinates A,B,C and D
\State \qquad ROIcircle=false(row, column)\Comment{Initialize logical circuit}

\qquad[x, y] = meshgrid(1:columns, 1:rows); 

\qquad$ROIcircle((x - A).^2 + (y - B).^2 <= Radius.^2) = true;$ 
\State $ROIsquare(A:B, C:D) = uint8(200)$\Comment{Initialize logical square}
\State I2=N\Comment{Initialize image}
\State $I_{2}(~ROIcircle)=1$\Comment{Zero image outside the circle mask}
\State $I_{2}(~ROIsquare)=1$\Comment{Zero image outside the square mask}
\EndProcedure
\end{algorithmic}
\end{algorithm}

\section{Results}
Results section covers performance evaluation of Variable Pixel G-Neighbor Filter and Hardware parameters.

\subsection{Variable Pixel G-Neighbor Filter vs Conventional Convolution Filter}
\begin{figure}[tbh]
\begin{centering}
\subfloat[\label{fig:1}]{\centering{}
\includegraphics[width=0.1\columnwidth]{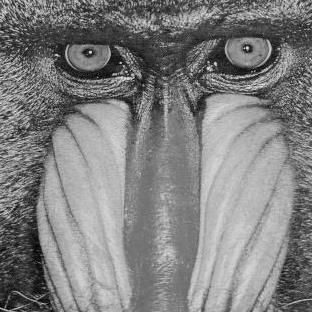}

}\quad \subfloat[\label{fig:2}]{\centering{}\includegraphics[width=0.1\columnwidth]{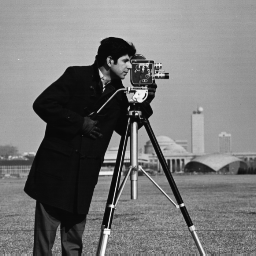}

}\quad \subfloat[\label{fig:3}]{\centering{}\includegraphics[width=0.1\columnwidth]{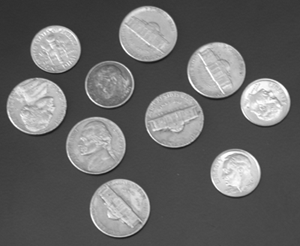}

}\quad \subfloat[\label{fig:4}]{\centering{}\includegraphics[width=0.1\columnwidth]{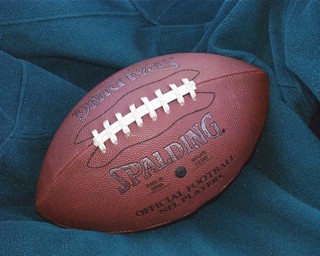}

}\quad \subfloat[\label{fig:5}]{\centering{}\includegraphics[width=0.1\columnwidth]{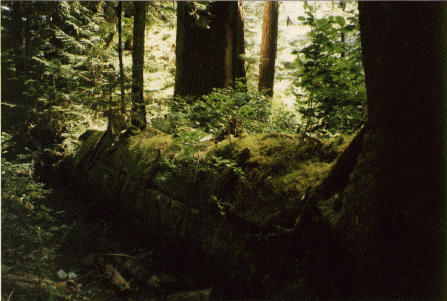}

}\quad \subfloat[\label{fig:6}]{\centering{}\includegraphics[width=0.1\columnwidth]{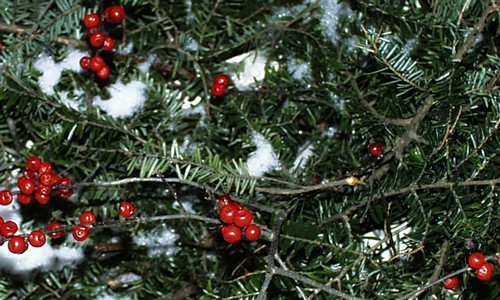}

}\quad \subfloat[\label{fig:7}]{\centering{}\includegraphics[width=0.1\columnwidth]{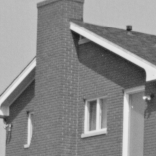}

}\quad \subfloat[\label{fig:8}]{\centering{}\includegraphics[width=0.1\columnwidth]{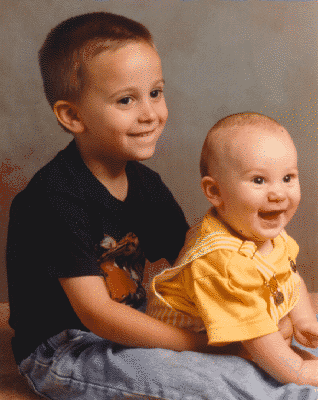}

}\quad \subfloat[\label{fig:9}]{\centering{}\includegraphics[width=0.1\columnwidth]{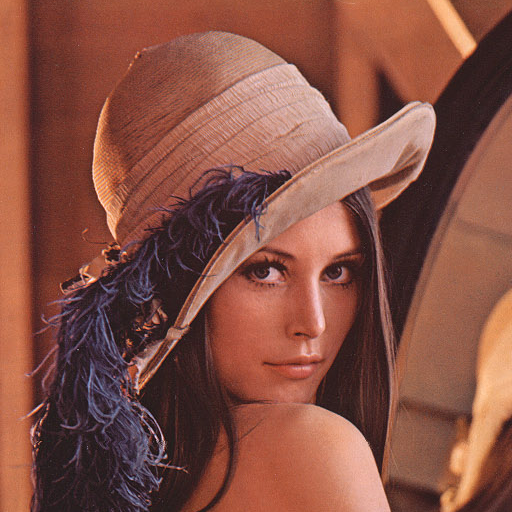}

}\quad \subfloat[\label{fig:10}]{\centering{}\includegraphics[width=0.1\columnwidth]{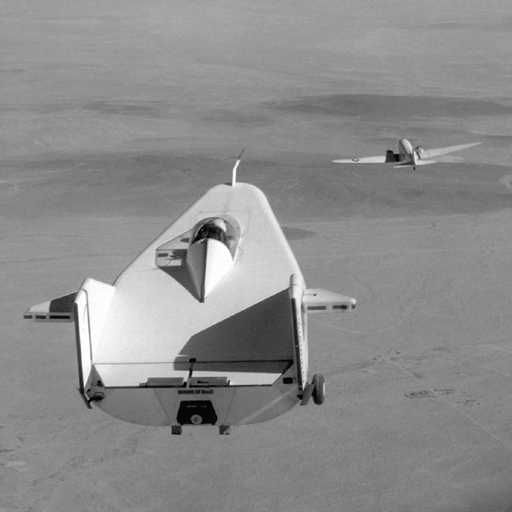}

}\quad \subfloat[\label{fig:11}]{\centering{}\includegraphics[width=0.1\columnwidth]{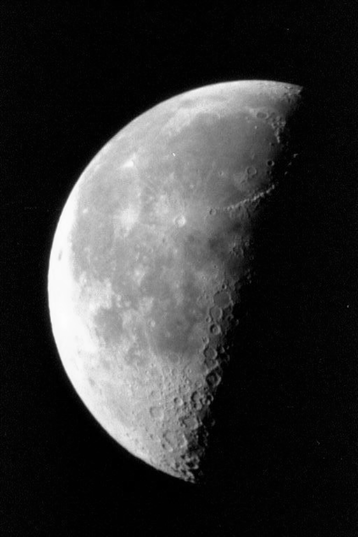}

}\quad \subfloat[\label{fig:12}]{\centering{}\includegraphics[width=0.1\columnwidth]{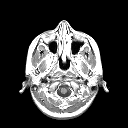}

}\quad \subfloat[\label{fig:13}]{\centering{}\includegraphics[width=0.1\columnwidth]{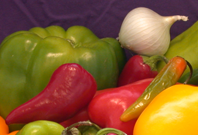}

}\quad \subfloat[\label{fig:14}]{\centering{}\includegraphics[width=0.1\columnwidth]{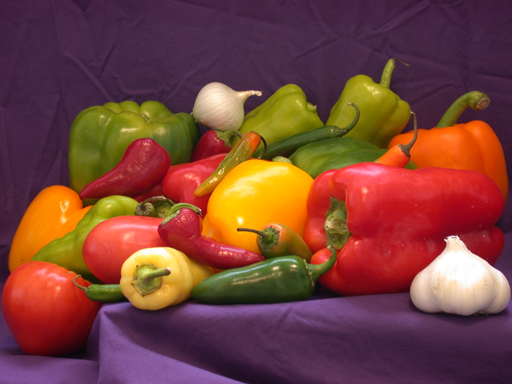}

}\quad \subfloat[\label{fig:15}]{\centering{}\includegraphics[width=0.1\columnwidth]{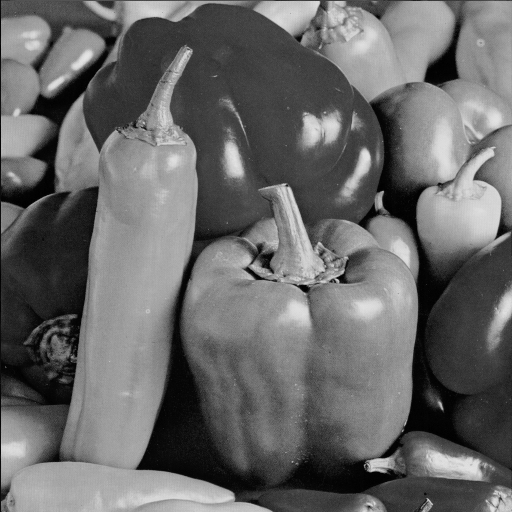}

}\quad \subfloat[\label{fig:16}]{\centering{}\includegraphics[width=0.1\columnwidth]{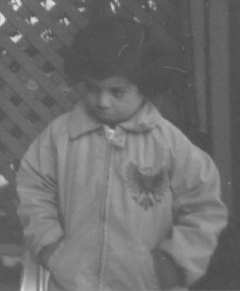}

}\quad \subfloat[\label{fig:17}]{\centering{}\includegraphics[width=0.1\columnwidth]{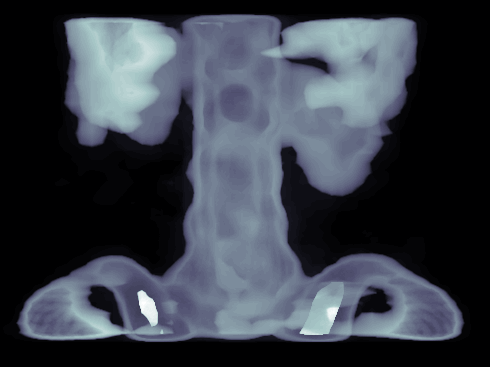}

}\quad \subfloat[\label{fig:18}]{\centering{}\includegraphics[width=0.1\columnwidth]{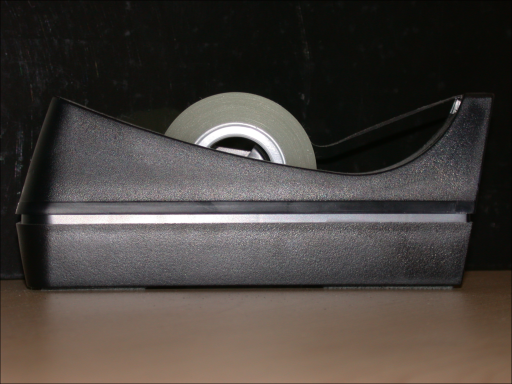}

}\quad \subfloat[\label{fig:19}]{\centering{}\includegraphics[width=0.1\columnwidth]{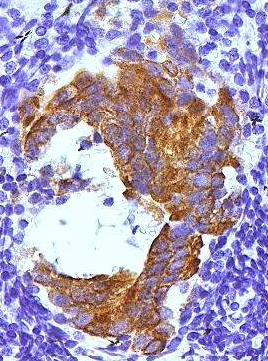}

}\quad \subfloat[\label{fig:20}]{\centering{}\includegraphics[width=0.1\columnwidth]{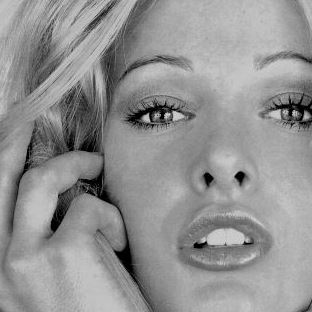}}
\par\end{centering}

\protect\caption{Set of standard images used}
\end{figure}

The simulation of proposed algorithm against conventional filters is implemented in MATLAB. Moreover, image performance metrics computation was also performed and numerically compared using both full reference and reduced reference approaches.
The set of standard images demonstrated in figures \ref{fig:1} to \ref{fig:20} representing 1-20 picture number in graphs correspondingly, was used to evaluate the effectiveness of the proposed method. The images were chosen to have gray and RGB nature and contain a variety of formats, including jpg and png. 

Firstly, seven images were separately investigated to find an optimal threshold value for the circuit. The threshold values in range from 0 to 0.3 were investigated on both PSNR and SSIM.\par

\begin{figure}[tbh]
\begin{centering}
\subfloat[\label{fig:original}]{\centering{}
\includegraphics[width=0.25\columnwidth]{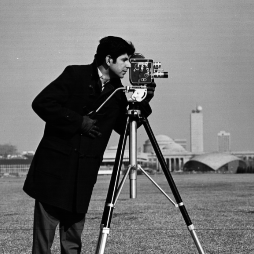}

}\quad \subfloat[\label{fig:mean}]{\centering{}\includegraphics[width=0.25\columnwidth]{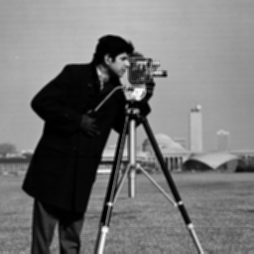}

}\quad \subfloat[\label{fig:meanG}]{\centering{}\includegraphics[width=0.25\columnwidth]{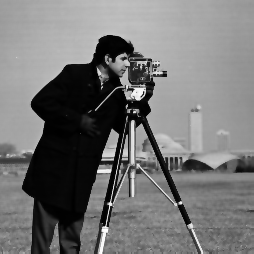}}
\par\end{centering}

\protect\caption{ Comparison of images  (a) original (b) G=0.3 (c) G=0.05}
\end{figure}

The PSNR memristive evaluation showed that starting from 0.3 till 0.1 PSNR values made a rise of about 3 dB. After this, the deviations of maximum results were in the range from 0.1 till 0.03. The threshold values below 0.03 were considered to be less effective. It can be deduced from PSNR and SSIM quality metrics, measured for different threshold values, that for memristive circuit optimal G is found to be 0.0507 for 0-1 range, or equivalently approximately 13 for 0-255 range. It can be subjectively observed from figure \ref{fig:mean} and \ref{fig:meanG} that obtained threshold value provides sufficient output.

\subsubsection{Full reference evaluation}

In this section, full reference method will be implemented using threshold obtained. Firstly, the demonstration of processing mean and median filtering operation on square window shape and adaptive window shape was done, see Fig. \ref{fig:comp}. It can be seen that application of adaptive filter without noise implementation leads to edge and structure preservation.

\begin{figure}[tbh]
\begin{centering}
\subfloat[\label{fig:original}]{\centering{}
\includegraphics[width=0.4\columnwidth]{cameraman.png}

}\quad \subfloat[\label{fig:noise}]{\centering{}\includegraphics[width=0.4\columnwidth]{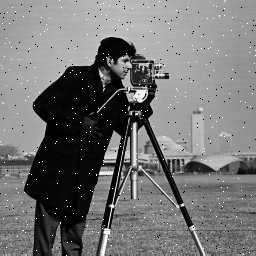}

}\quad \subfloat[\label{fig:mean1}]{\centering{}\includegraphics[width=0.4\columnwidth]{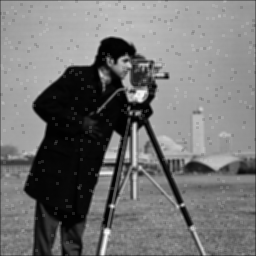}

}\quad \subfloat[\label{fig:Mmean}]{\centering{}\includegraphics[width=0.4\columnwidth]{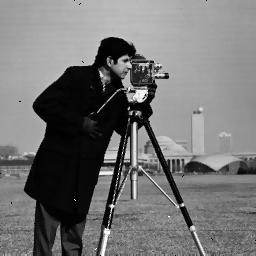}

}\quad \subfloat[\label{fig:median}]{\centering{}\includegraphics[width=0.4\columnwidth]{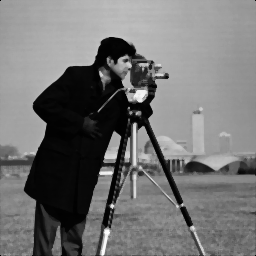}

}\quad \subfloat[\label{fig:Mmedian}]{\centering{}\includegraphics[width=0.4\columnwidth]{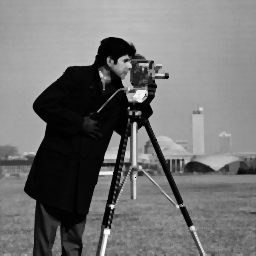}}
\par\end{centering}

\label{fig:comp}
\protect\caption{Full Reference evaluation of images a) original b) noise added c) mean square d) mean adaptive e) median square f) median adaptive  }\label{fig:comp}
\end{figure}

The salt and pepper noise was added to each image to assess the effectiveness of the proposed method. The data of image metrics with noise added obtained during image processing of mean and median filtering operation on square window shape and memristive pixel adaptive window shape collected in Table \ref{full}.

\begin{table}[tbh]
\centering
\caption{Full reference evaluation}
\label{full}
\begin{tabular}{llll}\hline
                & MSE            & PSNR        & SSIM           \\\hline
Mean
\\
Square     & 0.0027$\pm$ 0.0026 & 27.34$\pm$ 3.87 & 0.7794$\pm$ 0.0726 \\\hline
Mean 
\\
Adaptive   & 0.0009$\pm$ 0.0012 & 32.56$\pm$ 4.36 & 0.9373$\pm$ 0.0329 \\\hline
Median 
\\
Square   & 0.0017$\pm$ 0.0023 & 30.62$\pm$5.32  & 0.8717$\pm$ 0.0916 \\\hline
Median \\
Adaptive & 0.0017$\pm$ 0.0022 & 30.69$\pm$ 5.18 & 0.8770$\pm$ 0.0870\\\hline
\end{tabular}
\end{table}

\begin{figure*}[tbh]
\begin{centering}
\subfloat[\label{fig:MSE}]{\centering{}
\includegraphics[width=0.65\columnwidth]{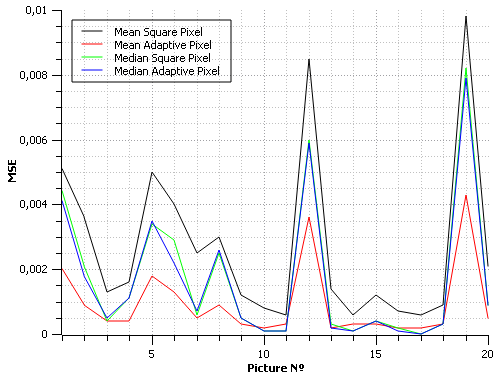}

}\quad \subfloat[\label{fig:PSNR}]{\centering{}\includegraphics[width=0.65\columnwidth]{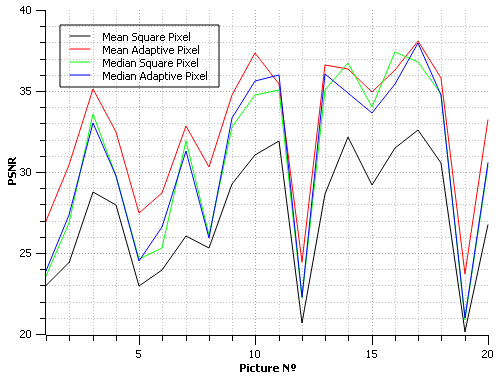}

}\quad \subfloat[\label{fig:SSIM}]{\centering{}\includegraphics[width=0.65\columnwidth]{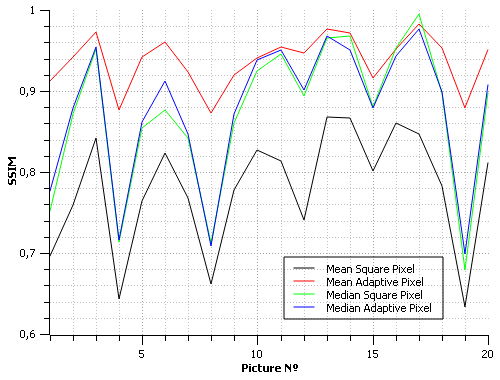}}
\par\end{centering}

\protect\caption{ Full reference evaluation of  (a) MSE (b) PSNR (c) SSIM}\label{fig:full}
\end{figure*}

It is clearly demonstrated in Fig. \ref{fig:full}, that both PSNR and SSIM of mean adaptive pixels outnumber conventional approaches and median adaptive pixels as well. 

\subsubsection{Reduced reference evaluation}

\begin{figure}[tbh]
\begin{centering}
\subfloat[\label{fig:1e}]{\centering{}
\includegraphics[width=0.4\columnwidth]{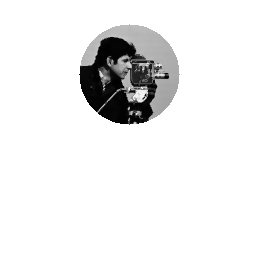}

}\quad \subfloat[\label{fig:1o}]{\centering{}\includegraphics[width=0.4\columnwidth]{cameraman.png}

}\quad \subfloat[\label{fig:9e}]{\centering{}\includegraphics[width=0.4\columnwidth]{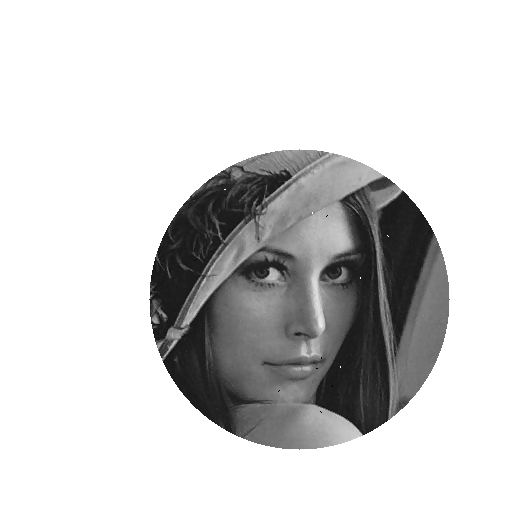}

}\quad \subfloat[\label{fig:9o}]{\centering{}\includegraphics[width=0.4\columnwidth]{lena.png}

}\quad \subfloat[\label{fig:20e}]{\centering{}\includegraphics[width=0.4\columnwidth]{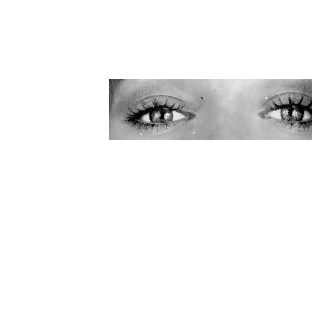}

}\quad \subfloat[\label{fig:20o}]{\centering{}\includegraphics[width=0.4\columnwidth]{woman_blonde.jpg}}
\par\end{centering}

\protect\caption{ Reduced Reference evaluation}\label{fig:rr}
\end{figure}

The reduced reference assessment was performed in the same manner as the full reference, the only difference that all measurements consider only the regions of interest (see Fig. \ref{fig:rr}).

\begin{figure*}[tbh]
\begin{centering}
\subfloat[\label{fig:MSERR}]{\centering{}
\includegraphics[width=0.65\columnwidth]{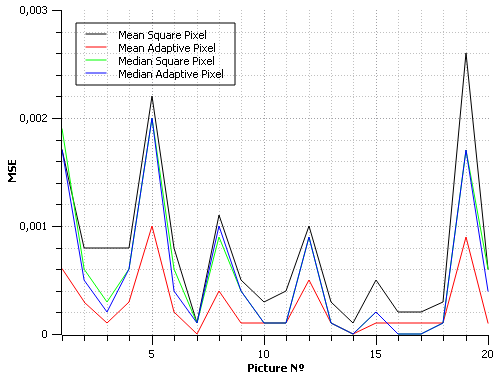}

}\quad \subfloat[\label{fig:PSNRRR}]{\centering{}\includegraphics[width=0.65\columnwidth]{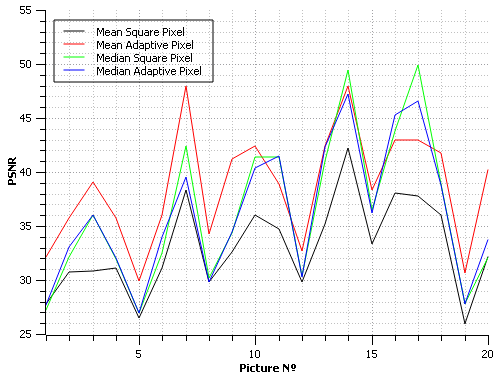}

}\quad \subfloat[\label{fig:SSIMRR}]{\centering{}\includegraphics[width=0.65\columnwidth]{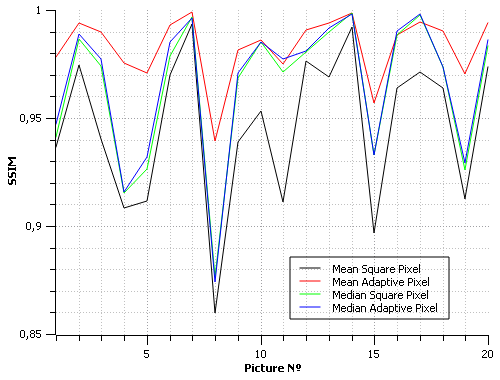}}
\par\end{centering}

\protect\caption{ Reduced reference evaluation of  (a) MSE (b) PSNR (c) SSIM}\label{fig:reduced}
\end{figure*}

The data collected in Table \ref{reduced} and visually demonstrated in Fig. \ref{fig:reduced} showed that PSNR improvement was moderate and SSIM indexes values also were improved.Median filtering reduced reference evaluation of the circuit followed the same trend as a full reference median evaluation. 

\begin{table}[tbh]
\centering
\caption{Reduced reference evaluation}
\label{reduced}
\begin{tabular}{llll}\hline
                & MSE            & PSNR        & SSIM           \\\hline
Mean \\
Square& 0.0008 $\pm$ 0.0007 & 32.99$\pm$ 4.27 & 0.9459$\pm$ 0.0358 \\\hline
Mean\\ 
Adaptive& 0.0003$\pm$ 0.0003 & 38.66$\pm$ 5.23 & 0.9830$\pm$ 0.0151 \\\hline
Median \\  
Square & 0.0006$\pm$ 0.0006 & 36.31$\pm$6.95  & 0.9646$\pm$ 0.0333 \\\hline
Median \\
Adaptive& 0.0005$\pm$ 0.0006 & 36.17$\pm$ 6.32 & 0.9665$\pm$ 0.0336\\\hline
\end{tabular}
\end{table}

Generally, the MATLAB simulation's results showed that proposed method of adaptive mean filtering provides a better performance in terms of full reference and reduced reference evaluation compared to conventional mean filter. 



\subsection{Hardware Parameters}
Memristive device in \cite{24} with threshold voltage of 1.088V and its SPICE model by Yakopcic \cite{25} was used. $R_{ON}$ and $R_{OFF}$ for this device are 0.125 M$\Omega$ and 1.14M$\Omega$, respectively. For switches and inverters, 180 nm CMOS technology from TSMC (Taiwan Semiconductor Manufacturing Company) was chosen due to its high compatibility with nanoscale memristors.  Simulations were carried out in Spice environment. \par
\begin{figure}[h]
\centering
\includegraphics[width=3in, height=1.5in]{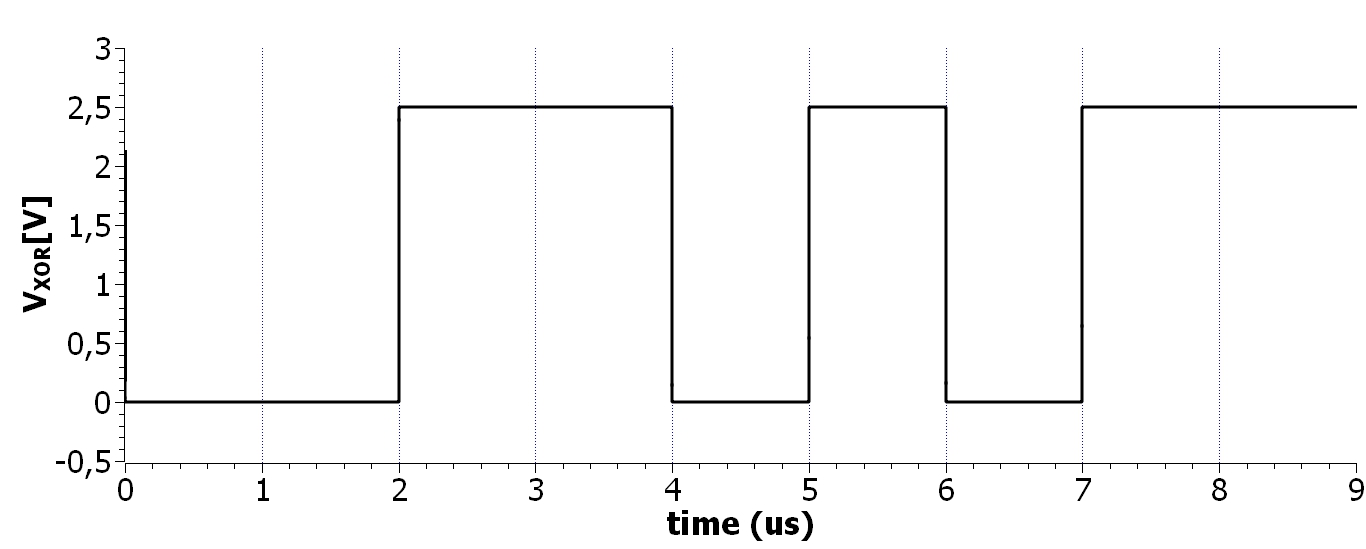}\\
\caption{XOR operation for inputs '11110011' and '11000110' }
\label{fig:xorresult}
\end{figure}
\begin{figure}[h]
\subfloat[\label{fig:sim}]{\centering{}
\includegraphics[width=3in]{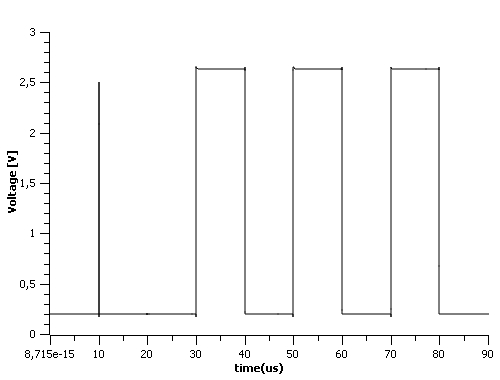}
}\quad \subfloat[\label{fig:sim1}]{\centering{}\includegraphics[width=3in]{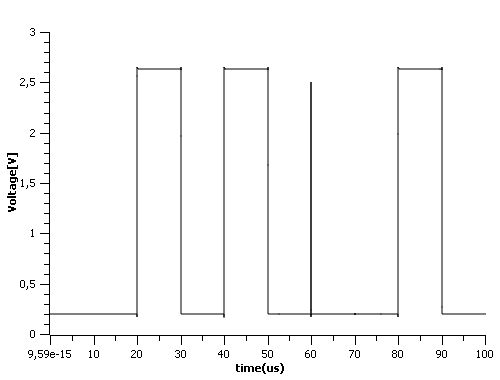}}
\caption{ Stored sequences in memristor array (a) '0111' and (b) '1101'\label{fig:similarity-g}}
\end{figure}
\begin{figure}[h]
\centering
\includegraphics[width=3in]{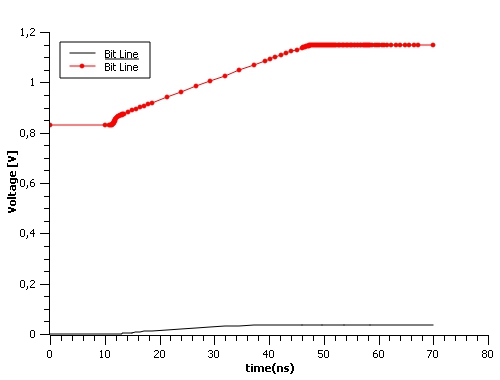}\\
\caption{Bit '1' stored in SRAM  }
\label{fig:sramoutput}
\end{figure}

$V_L$ and $V_H$ were selected to be 0 and 2.5V. Fig. \ref{fig:xorresult}  shows result of XOR operation of input waveforms $V_1$ = '11110011' and $V_2$ = '11000110'. The waveforms have duration of 1$\mu$s and duty cycle of 100\%. The resultant waveform is saved in one set of memristors, whereas threshold value is already in another set. For illustration purposes, four bits of distance and threshold values were compared.  In Fig. \ref{fig:similarity-g},  two bit sequences ('0111' and '1101') were read from the memristor array. One might notice that result of comparison of respective bits is '1' which is saved in SRAM as indicated in Fig. \ref{fig:sramoutput}.\par
 
Simulation of the proposed circuit demonstrates satisfactory operation. Use of memristor threshold logic to implement logic gates, such as AND, OR or XOR, allows to achieve reduced fabrication area. In addition to this, high packaging density is achieved using memristor crossbar for storage of intermediate bit values. Power and area calculations are illustrated in Table \ref{table_area_power}.\par
There are few works in literature on G-neighbor filtering and one of them focuses on hardware realization of this method using CMOS circuits \cite{yerbol}. The design employs current intensity values from image sensor and consists of separate stages such as max-min current selector, subtraction, exponential and threshold comparator circuits. Total power consumption and area of the circuit of two pixels were 911.5 mW and 303.5 ${\mu}m^{2}$ respectively. In comparison, hardware performance parameters of the memristive circuit based design   were 31.4 mW and 280.2 ${\mu}m^{2}$. 

\begin{table}[h]
\caption{Area and power consumption calculations for memristor-based G-neighbor identification circuit}
\label{table_area_power}
\begin{center}
    \begin{tabular}{lll}
    \hline 
    Blocks & Area (${\mu}m^{2}$) & Power ($mW$) \\ \hline
   XOR & 25.02 & 3.6\\ \hline
   Memristor arrays & 128.08 & 12\\ \hline
   Bit-by-bit comparison & 111.06 & 14 \\ \hline
   SRAM & 16 & 1.8    \\ \hline
   Total & 280.2 & 31.4    \\ \hline
     \end{tabular}
\end{center}
\end{table}

\section{Conclusion}
Given their high packaging density and CMOS compatibility, memristor has received much attention in recent years in applications such as non-volatile RAM, neuromorphic computing and pattern recognition. In this paper, memristor-based circuit is designed to perform G-neighborhood selection for image denoising applications. Simulation of the single unit has demonstrated the circuit can be employed to perform similarity identification between two pixels. 

\hl{}
\bibliographystyle{unsrt}
\bibliography{bibliography}

%





\end{document}